\documentclass[aps,prl,showpacs,twocolumn,superscriptaddress,citeautoscript]{revtex4}
\usepackage{graphics}
\usepackage{graphicx}
\usepackage{amsmath}
\usepackage{amsfonts}
\usepackage{epsfig}
\def\beq{\begin{equation}}
\def\eeq{\end{equation}}

\begin{document}

%Fill out the title section
\title{Multiplets Matter: The Electronic Structure of Rare-Earth Semiconductors and Semimetals}

%Author information
\author{Leonid~V.~Pourovskii}
\affiliation{Centre de Physique Th\'eorique, CNRS, \'Ecole Polytechnique, 91128 Palaiseau, France}
\author{Kris~T.~Delaney}
\affiliation{Materials Research Laboratory, University of California, Santa Barbara,
93106-5121, USA}
\author{Chris~G.~Van de Walle}
\affiliation{Materials Department, University of California, Santa Barbara,
93106-5050, USA}
\author{Nicola~A.~Spaldin}
\affiliation{Materials Department, University of California, Santa Barbara,
93106-5050, USA}
\author{Antoine~Georges}
\affiliation{Centre de Physique Th\'eorique, CNRS, \'Ecole Polytechnique, 91128 Palaiseau, France}

\begin{abstract}

We demonstrate that
%an {\it ab initio}
a theoretical framework fully incorporating intra-atomic
correlations and multiplet structure of the localized 4$f$ states is required in order
to capture the essential physics of rare-earth semiconductors and semimetals.
We focus in particular on the rare-earth semimetal erbium arsenide (ErAs),
for which effective one-electron approaches fail to
provide a consistent picture of both high and low-energy electronic states.
%In particular, we consider the case of the rare-earth semimetal erbium arsenide.
We treat the many-body states of the Er 4$f$ shell within an atomic
approximation in the framework of dynamical mean-field theory.
Our results for the magnetic-field dependence of the 4$f$ local moment,
the influence of multiplets on the photoemission spectrum, and
the exchange splitting of the Fermi surface pockets as measured from Shubnikov-de
Haas oscillations, are found to be in good agreement with experimental
results.
%
%% and compare its
%%predictions with those obtained by various one-electron
%%approaches as well as with experiment. The
%%Effective one-electron frameworks fail to
%%provide a coherent picture of both high and low-energy electronic states of ErAs.
%%In contrast,
%%The DMFT approach explains
%The observed behaviour of the 4$f$ local moment in applied
%high magnetic fields is explained by the interplay between the Zeeman and crystal-field
%splitting of the 4$f$ ground state multiplet. The shape of the Er 4$f$ quasiparticle
%bands is substantially modified due to multiplet effects,
%%and it is
%in good agreement with photoemission spectra.
%Interactions between the Er 4$f$ states and semi-metallic bands lead
%to exchange splitting of the , which is
%predicted by our approach in quantitative agreement with experiment.
\end{abstract}

\maketitle

The established framework for understanding the electronic structure of many materials
is based on independent electrons subject to a self-consistent potential.
This description often fails, however, when electronic states associated with
localized orbitals are involved.
Those states, while being too close in energy to the valence to be considered
as `core' states, do keep a strong atomic-like character.
For such materials, an entirely different point of view is required, in which intra-atomic
correlations and the full multiplet structure are taken into account from the start.
Here, we report on a particularly dramatic example: erbium arsenide, a semi-metal involving a
partially filled $4f$ shell.  Interest in this compound
was initially stimulated by growth, through molecular beam epitaxy, of ErAs onto pseudomorphically compatible
III-V semiconductor substrates of zinc-blende structure. The result is a
high-quality, epitaxial metallic contact with a continuous anion
sublattice through the interface\cite{Palmstrom/Tabatabaie/Allen:1988}.
Other growth
modes\cite{Palmstrom/Tabatabaie/Allen:1988,Singer_et_al:1994,Klenov_et_al:2005,Schultz/Palmstrom:2006}
have been found to yield metallic ErAs nanoparticles epitaxially embedded in the
semiconductor matrix, with
%Potential applications of such structures include
potential applications including
efficient thermoelectrics\cite{Kim/Zide_et_al:2006} and solid-state THz
emitters\cite{Kadow_et_al:2000}.

The low-energy electronic states of erbium arsenide (ErAs)
consist\cite{Palmstrom/Tabatabaie/Allen:1988}
of a conduction band and a valence band, with predominant Er $5d$ and
As $4p$ character respectively. The small energy-overlap of these bands is responsible
for the semimetallic nature of this material.
In contrast, the Er $4f$ electrons are well localized and form local moments.
Treating simultaneously these two kinds of electronic states is a major challenge for
conventional electronic structure theories.
Indeed, when the $4f$ states are treated as core electrons, calculations based
on the local-density approximation (LDA) do yield
%a semi-metallic band-structure,
a semi-metal, but do not otherwise provide an accurate description of the electronic properties.
The volume of the calculated FS pockets is approximately three times too large,
leading to incorrect carrier concentrations and SdH
frequencies\cite{Petukhov/Lambrecht/Segall:1994,Petukhov/Lambrecht/Segall:1996} compared
with accurate measurements\cite{Allen:1991,Bogaerts:1992,Bogaerts:1993,Bogaerts_et_al:1996}.

Here we show that a proper treatment of the strong correlations and multiplet
structure of the $4f$ shell provides a solution to these difficulties.
Our finding is general and has fundamental consequences for
the understanding of the electronic structure of a large class of
rare-earth compounds. ErAs is a particularly remarkable example, however, in
which the electronic states {\it near the Fermi level} act as a sensitive probe of the
atomic physics associated with more strongly bound localized states.

Our approach is based on the dynamical mean-field theory\cite{geo96} (DMFT), combined\cite{voll05,kotliar_review} with LDA.
The $4f$ shell is treated as that of an effective atom self-consistently coupled to an
environment describing the rest of the solid.
From the hamiltonian of this effective atom (which takes into account crystal-field
effects, intra-atomic Coulomb interactions and the spin-orbit coupling),
a many-body self-energy is computed within the Hubbard-I approximation \cite{hubbard_1} and inserted
into the Green's function of the full solid. Self-consistency over
the total charge density and the effective atom parameters is implemented. The full description of our approach
can be found in Ref.~\cite{SCDMFT}. In the present work  it was generalized to include the spin-orbit interaction as well
as the full 4-index local Coulomb interaction matrix.  The parameter $U=7.94$ eV  of the local Coulomb interaction on the Er 4$f$ shell has been determined by constrained LDA calculations, while the Slater integrals $F^2=12.1$~eV, $F^4=8.4$~eV, and $F^6=6.7$~eV,  which are known to be weakly dependent on the crystalline environment, have been taken from the optical measurements on Er ions embedded in a LaF$_3$ host \cite{Carnall89}. To perform LDA+U calculations we have used the same framework, including
the spin-orbit interaction, with the Hubbard-I self-energy being substituted by a local self-energy computed within the rotationally-invariant Hartree-Fock expression.
All our calculations have been carried out in the rock-salt structure at the ErAs experimental lattice parameter of 5.74 \AA.

%
%In order to make contact with SdH and magnetoresistance experiments, we also consider
%the effect of an applied magnetic field $H$ (coupling to $2\mathbf{S}+\mathbf{L}$).

\begin{figure}[t]
\begin{center}
\includegraphics[width=0.85\columnwidth]{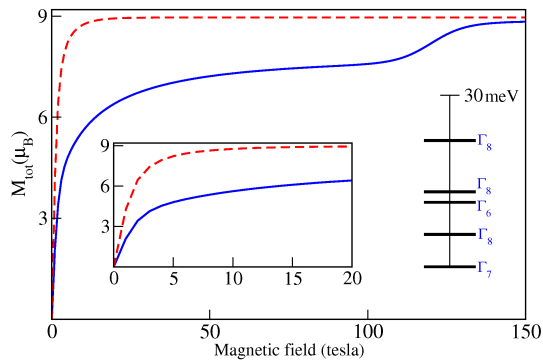}
\end{center}
\caption{\label{fig:magnetization} (Color online)
Calculated DMFT total magnetic moment on Er 4$f$ shell vs. applied magnetic field
at $T=4.2$\,K with (solid blue curve) and without (red dashed curve)
crystal field effects taken into account. The left inset shows
the zoom in the field's range up to $20$\,T. The crystal field splitting of
the $^4I_{15/2}$  multiplet obtained by LDA+DMFT in paramagnetic ErAs
is shown in the right inset.
}
\end{figure}

The ground-state of an isolated Er$^{3+}$ ion ($4f^{11}5d^0$) is a 16-fold degenerate
multiplet $^4I_{15/2}$ ($S=\frac{3}{2},L=6,J=\frac{15}{2}$).
In a cubic crystal-field and at zero magnetic field, this multiplet is expected to
split into twofold degenerate $\Gamma_6$ and $\Gamma_7$ multiplets, as well as three
$4$-fold degenerate $\Gamma_8$ multiplets \cite{Lea1962}. At self-consistency, the zero-field eigenstates of
our effective atom hamiltonian (inset of Fig.~\ref{fig:magnetization}) follow this
expectation, the obtained ground-state being the $\Gamma_7$ multiplet.
%
%with a total moment of $\pm2.86\mu_B$ for spin up/down.
%
As a magnetic field $H$ is turned on, the ground-state
remains approximately $\Gamma_7$ for a wide range of field, and can be decomposed
on eigenstates of $J_z$ according to:
$a_{13/2}^{(H)}|J_z=\frac{13}{2}\rangle+a_{5/2}^{(H)}|J_z=\frac{5}{2}\rangle+
a_{-3/2}^{(H)}|J_z=-\frac{3}{2}\rangle+a_{-11/2}^{(H)}|J_z=-\frac{11}{2}\rangle$
, where the coefficients $a_{J_z}$ depend on magnetic field.
%AG
% Comment on low-T, in relation to Neel temp ?
%
The magnetic moment {\it vs.} field curve (Fig.~\ref{fig:magnetization}) displays an
initial sharp rise due to the Zeeman splitting of the $\Gamma_7$ state,
followed by a rather slow increase in the range from $5$ to $120$T due to the progressive
polarization of the $\Gamma_7$ state.
From magnetoresistance experiments\cite{Bogaerts_et_al:1996}, it was
proposed that the magnetic moment is saturated at a value\cite{Allen_et_al:1989} $\sim 5.3\mu_B$
in fields above $10\rm{T}$.
Indeed, we find values ranging from $4.8\mu_B$ at $H=5\rm{T}$ to $5.6\mu_B$ at
$H=10\rm{T}$, in good agreement with the experimental estimate\cite{Allen_et_al:1989}.
However, our analysis shows that this is rather a quasi-saturation, with the moment being ``frozen''
by the crystal field. Eventually, at a very high field $H\simeq 120 T$ we predict a sharp increase of
the moment to $8.9\mu_B$, corresponding to a symmetry-changing transition of the ground-state
from $\Gamma_7$ to $\Gamma_8$.
%
%The composition of the $\Gamma_7$ crystal field state can be expressed  through the many-body
%eigenstate of the total momentum operator as $a_1\ket{m_J=13/2} +a_2 \ket
%{m_J=5/2}+a_3\ket{m_J=-3/2}+a_4\ket{m_J=-11/2}$ \cite{Lea1962}. The coefficients
% $a$ change rather slowly as function of the field, moreover, for $\Gamma_7$ the
%highest moment reachable corresponds to that of the  $\ket{m_J=13/2}$ state, 7.8 $\mu_B$,
%still below the Hund's rule value of 9 $\mu_B$.
%

It is interesting to contrast these findings to those of LDA+U calculations.
We performed these for ferromagnetic ordering since, for fields above $\sim 1\rm{T}$
relevant to SdH experiments, this is the observed configuration\cite{Allen_et_al:1989}.
In the LDA+U approach \cite{anisimov_lda+u_review_1997_jpcm}, Coulomb interaction effects within the $4f$-shell
are described by a self-consistent, orbital- and spin- dependent, {\it one-electron potential}.
As a result, the LDA+U solution of lowest energy is found to correspond to the filling of
independent electron levels according to Hund's rules\cite{Larson2007}, leading to the highest possible
spin and orbital moments consistent with a $4f^{11}$ shell.
Indeed, our LDA+U results yield a spin, orbital and total moment of $2.9\mu_B$, $6.0\mu_B$,
and $8.9\mu_B$ respectively, in contradiction with the observed experimental value quoted
above.
%
% AG Comment on large-field solution recovering LDA+U ?
%
Furthermore, as pointed out by Larson {\it et al.} for nitrides~\cite{Larson2007},
the LDA+U solution breaks the cubic symmetry of the lattice,
because it is constructed by occupying one-electron eigenstates of the angular momentum operator
$\hat{L}_z$. LDA+U solutions preserving cubic symmetry are obtained by occupying
symmetry-adapted one-electron states (which are {\it not} eigenstates of $\hat{L}_z$), but they
have higher energy.
Hence, in a one-electron effective description, the system has to choose between minimizing the
intra-atomic Coulomb interaction energy and preserving the lattice symmetry. In reality however,
this dilemma does not apply since atomic multiplets satisfy both requirements. A proper many-body
treatment of atomic correlations avoids this conundrum, as demonstrated above.
\begin{figure*}
\begin{center}
\includegraphics[width=1.9\columnwidth]{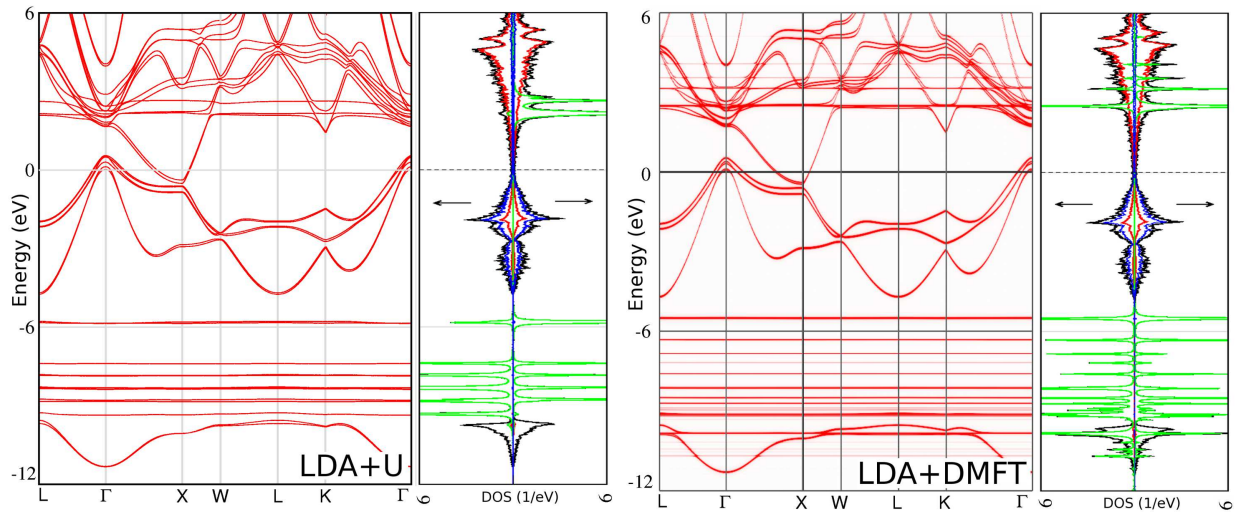}
\end{center}
\caption{\label{fig:bandstruc}
Band structure and density of states (DOS) of ErAs:
from LDA+U (left), and LDA+DMFT at an applied field of $5$\,T (right).
In the corresponding DOS plots the total, Er $5d$, Er $4f$ and As $4p$ states
are displayed by the black, red, green, and blue curves respectively.
}
\end{figure*}
We now describe the electronic structure of ErAs, obtained with
LDA+U and LDA+DMFT (Fig.~\ref{fig:bandstruc}).
Distinctive features corresponding to the $4f$ states are observed at high energy,
corresponding to the upper- and lower- Hubbard bands (UHB/LHB).
In LDA+DMFT, those ``bands'' truly are many-body atomic-like excitations
associated with the removal (LHB) or addition (UHB) of an electron in the $4f$ shell.
The LHB spans an
energy range from $\sim -11$\, eV to $-5.5$\, eV,
in agreement with photoemission experiments\cite{Komesu_et_al:2003,Duan2004}.
%
%The main peak forming the UHB within LDA+DMFT is found at $2-3\,\rm{eV}$ above the Fermi energy,
% with additional multiplet peaks at $\sim 3-4\,\rm{eV}$.
%Hence, the UHB spans an energy range of $\sim 1.5\,\rm{eV}$, which is consistent with its
%apparent width as measured in inverse photoemission experiments\cite{Komesu_et_al:2003,Duan2004}.
Within LDA+DMFT the UHB consists not only of the main peak at $2-3\,\rm{eV}$ above the Fermi energy, but also
 of additional multiplet peaks at $\sim 3-4\,\rm{eV}$. As also discussed for $\delta$-Pu\cite{shim07},
this multiplet structure is responsible for the apparent width of the UHB, which we find to be $\sim 1.5\,\rm{eV}$,
in agreement with inverse photoemmission measurements \cite{Komesu_et_al:2003,Duan2004}.
% which cause a broadening of the UHB to $\sim 1.5\,\rm{eV}$.
%This value of the UHB's width agrees well with inverse photoemission measurements of Refs.~\cite{Komesu_et_al:2003,Duan2004}.
%A broadening of the Hubbard bands due to multiplet effects has been previously discussed\cite{shim07} for the case of $\delta$-Pu.

%
In contrast, the UHB found with LDA+U spans an energy range of only $\sim 0.75\,\rm{eV}$,
because the additional multiplet peaks are missed in this approach.
Furthermore, a distinctive feature found within LDA+U is that the UHB is
almost fully (minority) spin-polarized. This is clearly due to the maximal
(Hund's rule) spin moment, corresponding to the complete
filling of the $4f$ majority-spin states so that no electron addition
is possible in the majority channel.
In contrast, the UHB found within LDA+DMFT is found to have only partial spin
polarization. Hence UHB transitions
are possible in both spin channels, a distinctive prediction for possible spin-polarized
photoemission experiments.

\begin{figure*}
\begin{center}
\includegraphics[width=1.9\columnwidth]{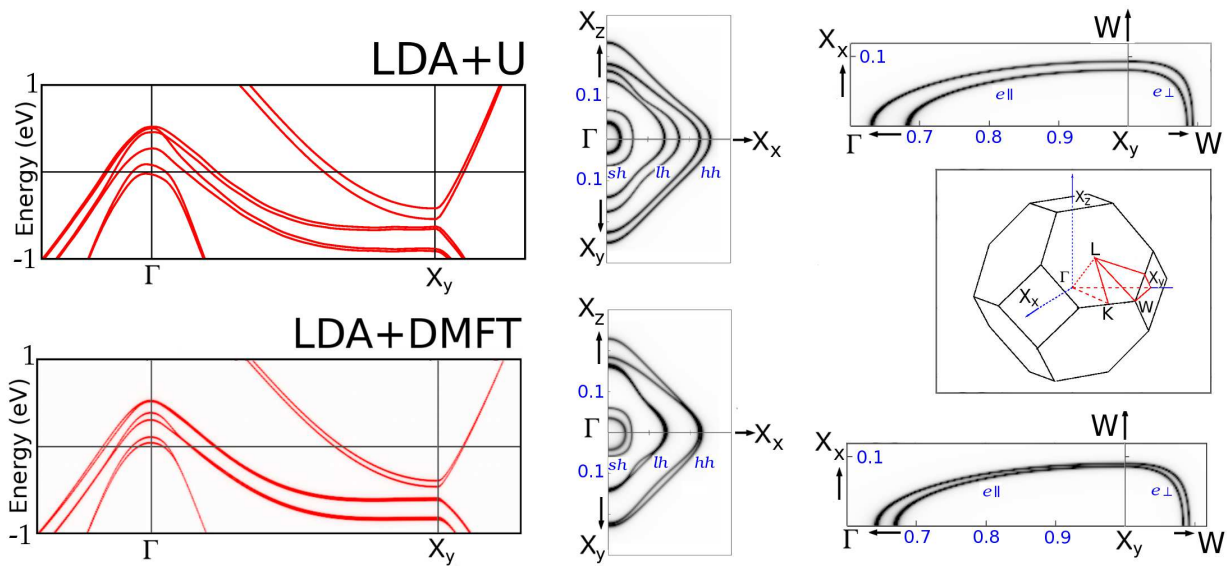}
\end{center}
\caption{\label{fig:FS} (Color online)
The low-energy electronic structure of ErAs within the LDA+U(top) and LDA+DMFT (bottom) approaches.
 The spin polarization is along the $z$ axis,
thus the cubic symmetry is lifted, and the band structure along $x(y)$ and $z$ axis are not equivalent.
Therefore we designate the corresponding $X$ high-symmetry points as $X_x$, $X_y$, and $X_z$.
The Brillouin zone is show in insert at the right-hand side of the plot.
The overlapping As 4$p$ and Er 5$d$ bands near E$_F$ are shown on the left-hand side.
They are followed to the right by the cross sections of the hole pockets at the $\Gamma$ point
in the $xy$ and $xz$ planes, and the longitudinal $e\parallel$ (in the $\Gamma X_yW$ plane) and
transverse $e\perp$ (in the $X_yWW$ plane) cross sections of the electronic pocket.
The exchange splittings of the pockets are clearly seen.
}
\end{figure*}

We now show that these differences in the description of the high-energy $4f$
states have key consequences for the electronic properties of states near the Fermi level.
%
%The low-energy band-structure and Fermi surface (FS) of ErAs is highlighted in Fig.~\ref{fig:FS}.
%It consists in As $4p$ and Er $5d$ bands, with a small overlap responsible for the
%semi-metallic character of this material.
%
Near the $\Gamma$ high-symmetry point, the As $4p$ bands form ``heavy-'' ($hh$) and ``light-'' ($lh$)
hole pockets, as well as a small hole pocket ($sh$) due to the spin-orbit splitting of the $4p$ states,
while the ellipsoidal electronic pocket at the $X$-point is associated with the Er 5$d$- states (Fig.~\ref{fig:FS}).
\begin{table}
\
\caption{\label{supl_table} Shubnikov-de Haas(SdH) frequencies (in tesla)and carrier concentrations of electrons/holes $n$ (in 10$^{20}\rm{cm}^{-3}$). The frequencies $f_{sh}$, $f_{lh}$, and $f_{hh}$ correspond to cross sections of the small, light, and heavy hole pockets near the $\Gamma$ high-symmetry point, respectively. The frequencies $e\perp$ and $e\parallel$ are due to the transverse and longitudinal cross sections of the ellipsoidal electronic pocket. The experimental carrier concentrations and SdH frequencies are for an Er$_{0.57}$Sc$_{0.43}$As alloy \cite{Bogaerts_et_al:1996}.}
\begin{ruledtabular}
\begin{tabular}{l|cc|c|cc|cc}
Orbits & \multicolumn{2}{c|}{Exp.}& LDA  & \multicolumn{2}{c|}{LSDA+U}  & \multicolumn{2}{c}{LDA+DMFT} \\
 & $\uparrow$ &  $\downarrow$ & $\uparrow$/$\downarrow$   & $\uparrow$ &  $\downarrow$  & $\uparrow$ &  $\downarrow$ \\
\hline
$f_{sh}$ & 150& 150 & 511 & 163 & 40 & 140 & 72 \\
$f_{lh}$ & 612& 589 & 1590 & 907 & 597 & 619 & 574 \\
$f_{hh}$ & 1273& 1222 & 2479& 1592 & 1907 & 1637 & 1466 \\
$f_{e\perp}$& 386& 328 & 479& 333 & 243  & 362 & 306 \\
$f_{e\parallel}$& 1111& 941 & 1848& 1205 & 850 &1270 &1113\\
\hline
$n$ & \multicolumn{2}{c|}{3.3}& 7.6 & \multicolumn{2}{c|}{4.1} & \multicolumn{2}{c}{3.9} \\
\end{tabular}
\end{ruledtabular}
\end{table}
%
%
%
% Mention SdH experiments ?
%
%As mentioned above, a conventional LDA approach treating the $4f$ shell as core states
%severely overestimates the $p$-$d$ overlap, and hence the carrier concentration
%($\sim 7.58\times 10^{20} \rm{cm}^{-3}$, in contrast to the experimental value
%$\sim 3.3\times 10^{20} \rm{cm}^{-3}$), as well as the size of the FS pockets (with
%SdH frequencies larger than experiments by approximately a factor of two, see supplementary
%information for a detailed table).
%%
%Both the LDA+DMFT and LDA+U treatment of the $4f$ states lead to a drastic reduction of
%both the carrier concentration ($\sim 4.1\times 10^{20} \rm{cm}^{-3}$ for LDA+U and
%$\sim 3.9\times 10^{20} \rm{cm}^{-3}$ for LDA+DMFT in a field of $H=5\rm{T}$), as well as of
%the size of the FS pockets (see supplementary table), in much better agreement with
%experiments.
%
As mentioned above, a conventional LDA approach treating the $4f$ shell as core states
severely overestimates the $p$-$d$ overlap, and hence the carrier concentration
and the size of the FS pockets.
Both the LDA+DMFT and LDA+U treatment of the $4f$ states lead to a drastic
 reduction in the
magnitude of the carrier density and SdH frequencies
(Table~\ref{supl_table}), and hence to a much better agreement with experiments.
%
%The carrier density, for example, is $7.58\times 10^{20} \rm{cm}^{-3}$ in our
%LDA ($f$ in core) calculations, while it is
%$4.1\times 10^{20} \rm{cm}^{-3}$ for LDA+U and
%$3.9\times 10^{20} \rm{cm}^{-3}$ for LDA+DMFT in an applied field $H=5\rm{T}$,
%to be compared with the experimental value\cite{Bogaerts_et_al:1996} $\sim 3.3\times 10^{20} \rm{cm}^{-3}$.
%
%For detailed values of the SdH frequencies of all FS pockets (which are roughly
%overestimated by a factor of two within LDA) see the Supplementary Table.
%
These results demonstrate that taking into account interaction effects in
the $4f$ shell is key to a proper description of the {\it valence} band structure
of ErAs.
% (of mostly $p$- and $d$-character).
%
Previous works employed an empirical shift of the $5d$ band
in order to correct for the inadequacies of LDA\cite{Petukhov/Lambrecht/Segall:1996,Lambrecht_et_al:1997}.
\begin{table}
%[!H]
\caption{\label{table:sdh} Exchange splittings of Shubnikov-de Haas frequencies
$\Delta=f^\uparrow-f^\downarrow$ (in tesla)}
\label{table}
\begin{ruledtabular}
\begin{tabular}{lcccc}
Orbits & Experiment\tablenote{Experimental data for an Er$_{0.57}$Sc$_{0.43}$As alloy\cite{Bogaerts_et_al:1996}.}  & LSDA+U  & GW\tablenote{Chantis {\it et al.}\cite{chantis07}}& LDA+DMFT \\
\hline
$\Delta_{lh}$& 23 & 310 & 136 & 45 \\
$\Delta_{hh}$& 51 & 315 & 274 & 171 \\
$\Delta_{e\perp}$& 58 & 90 & 151 & 56 \\
$\Delta_{e\parallel}$& 170 & 355 & 430 & 157 \\

\end{tabular}
\end{ruledtabular}
\end{table}
Despite the overall reduction in the FS pocket volume, the LDA+U (self-consistent one-electron potential) and
LDA+DMFT (self-consistently embedded atom) results differ significantly when it comes to
the exchange splitting of these pockets.
In Table~\ref{table:sdh}, we display the measured splittings\cite{Allen:1991,Bogaerts:1993,Bogaerts_et_al:1996} in SdH experiments
along with the calculated values from LDA+DMFT, LDA+U and a recent spin-polarized GW
work\cite{chantis07}.
It is apparent that LDA+U (and spin-polarized GW)
strongly overestimates the splitting of all orbits, especially for the hole
pockets, while a much better agreement with experiments is obtained from LDA+DMFT.
The reasons for this success are: i) a proper description of the crystal-field induced
freezing of the $4f$ spin moment (in contrast to the full polarization obtained from
LDA+U) and ii) a proper description of the orbital angular momentum content of the upper
Hubbard band. Electronic pockets are mainly sensitive to the
first effect, since the direct hybridization between the Er $d$- and $f$- states
is very small near the X-point\cite{Petukhov/Lambrecht/Segall:1996}.
As a result, the splitting of the electronic pocket is mainly due to the
exchange field induced by the polarization of the $f$-states in the local charge
density, and hence in the effective potential seen by the $d$-states \cite{Wulff88}. Because
the spin polarization of the $f$-states is strongly overestimated in LDA+U, so
is the exchange splitting of the electron pockets. In contrast, the hybridization
between the As $p$-states and Er $f$-states near the $\Gamma$-point is rather strong.
In LDA+U, the UHB has a strong spin-polarization in the minority channel: hybridization
thus repels the corresponding $p$-states away from the Fermi level, leading to too
small hole pockets in the minority channel (see Table~\ref{supl_table}). In
LDA+DMFT, this spin polarization is smaller and the orbital angular momentum
composition of the UHB is different,
with larger contributions from components which do not
hybridize with $p$-states for symmetry reasons. Hence the smaller exchange
splitting of the hole-pockets, in better agreement with experiments.
It is remarkable that, because of the sensitivity of hybridization effects to
both the spin and orbital angular momentum of the $4f$ states, experimental
measurements of the exchange splitting of the FS pockets indirectly probe the
electronic state of the Er $4f$ shell.

In conclusion, we have shown that a proper description of the electronic
structure of ErAs requires a conceptual framework which gives a central role to
local atomic physics and multiplet effects, in
contrast to conventional frameworks based on one-electron effective descriptions.
This is of general importance for a wide class of materials,
including all rare-earth based semiconductors and semimetals.

This work was supported by the Agence Nationale de la Recherche (France) under the ETSF Award, and by
the MRSEC Program of the National Science Foundation under Award No. DMR05-20415.
AG and LP thank the Chemical Bonding Center and the Kavli Institute for
Theoretical Physics, UCSB, for hospitality and support. LP acknowledges the financial support from ICAM
under NSF grant DMR 0645461.

%\bibliography{ErAs,bibextra,bibag}

\begin{thebibliography}{28}
\expandafter\ifx\csname natexlab\endcsname\relax\def\natexlab#1{#1}\fi
\expandafter\ifx\csname bibnamefont\endcsname\relax
  \def\bibnamefont#1{#1}\fi
\expandafter\ifx\csname bibfnamefont\endcsname\relax
  \def\bibfnamefont#1{#1}\fi
\expandafter\ifx\csname citenamefont\endcsname\relax
  \def\citenamefont#1{#1}\fi
\expandafter\ifx\csname url\endcsname\relax
  \def\url#1{\texttt{#1}}\fi
\expandafter\ifx\csname urlprefix\endcsname\relax\def\urlprefix{URL }\fi
\providecommand{\bibinfo}[2]{#2}
\providecommand{\eprint}[2][]{\url{#2}}

\bibitem[{\citenamefont{Palmstr{\o}m et~al.}(1988)\citenamefont{Palmstr{\o}m,
  Tabatabaie, and {Allen Jr.}}}]{Palmstrom/Tabatabaie/Allen:1988}
\bibinfo{author}{\bibfnamefont{C.~J.} \bibnamefont{Palmstr{\o}m}},
  \bibinfo{author}{\bibfnamefont{N.}~\bibnamefont{Tabatabaie}},
  \bibnamefont{and} \bibinfo{author}{\bibfnamefont{S.~J.} \bibnamefont{{Allen
  Jr.}}}, \bibinfo{journal}{Appl. Phys. Lett.} \textbf{\bibinfo{volume}{53}},
  \bibinfo{pages}{2608} (\bibinfo{year}{1988}).

\bibitem[{\citenamefont{Singer~\emph{et al.}}(1994)}]{Singer_et_al:1994}
\bibinfo{author}{\bibfnamefont{K.~E.} \bibnamefont{Singer~\emph{et al.}}},
  \bibinfo{journal}{Appl. Phys. Lett.} \textbf{\bibinfo{volume}{64}},
  \bibinfo{pages}{707} (\bibinfo{year}{1994}).

\bibitem[{\citenamefont{Klenov~\emph{et al.}}(2005)}]{Klenov_et_al:2005}
\bibinfo{author}{\bibfnamefont{D.~O.} \bibnamefont{Klenov~\emph{et al.}}},
  \bibinfo{journal}{Appl. Phys. Lett.} \textbf{\bibinfo{volume}{83}},
  \bibinfo{pages}{111912} (\bibinfo{year}{2005}).

\bibitem[{\citenamefont{Schultz and
  Palmstr{\o}m}(2006)}]{Schultz/Palmstrom:2006}
\bibinfo{author}{\bibfnamefont{B.~D.} \bibnamefont{Schultz}} \bibnamefont{and}
  \bibinfo{author}{\bibfnamefont{C.~J.} \bibnamefont{Palmstr{\o}m}},
  \bibinfo{journal}{Phys. Rev. B} \textbf{\bibinfo{volume}{73}},
  \bibinfo{pages}{241407(R)} (\bibinfo{year}{2006}).

\bibitem[{\citenamefont{Kim~\emph{et al.}}(2006)}]{Kim/Zide_et_al:2006}
\bibinfo{author}{\bibfnamefont{W.}~\bibnamefont{Kim~\emph{et al.}}},
  \bibinfo{journal}{Phys. Rev. Lett.} \textbf{\bibinfo{volume}{96}},
  \bibinfo{pages}{045901} (\bibinfo{year}{2006}).

\bibitem[{\citenamefont{Kadow~\emph{et al.}}(2000)}]{Kadow_et_al:2000}
\bibinfo{author}{\bibfnamefont{C.}~\bibnamefont{Kadow~\emph{et al.}}},
  \bibinfo{journal}{Appl. Phys. Lett.} \textbf{\bibinfo{volume}{76}},
  \bibinfo{pages}{3510} (\bibinfo{year}{2000}).

\bibitem[{\citenamefont{Petukhov et~al.}(1994)\citenamefont{Petukhov,
  Lambrecht, and Segall}}]{Petukhov/Lambrecht/Segall:1994}
\bibinfo{author}{\bibfnamefont{A.~G.} \bibnamefont{Petukhov}},
  \bibinfo{author}{\bibfnamefont{W.~R.~L.} \bibnamefont{Lambrecht}},
  \bibnamefont{and} \bibinfo{author}{\bibfnamefont{B.}~\bibnamefont{Segall}},
  \bibinfo{journal}{Phys. Rev. B} \textbf{\bibinfo{volume}{50}},
  \bibinfo{pages}{7800} (\bibinfo{year}{1994}).

\bibitem[{\citenamefont{Petukhov et~al.}(1996)\citenamefont{Petukhov,
  Lambrecht, and Segall}}]{Petukhov/Lambrecht/Segall:1996}
\bibinfo{author}{\bibfnamefont{A.~G.} \bibnamefont{Petukhov}},
  \bibinfo{author}{\bibfnamefont{W.~R.~L.} \bibnamefont{Lambrecht}},
  \bibnamefont{and} \bibinfo{author}{\bibfnamefont{B.}~\bibnamefont{Segall}},
  \bibinfo{journal}{Phys. Rev. B} \textbf{\bibinfo{volume}{53}},
  \bibinfo{pages}{4324} (\bibinfo{year}{1996}).

\bibitem[{\citenamefont{Allen Jr.~\emph{et al.}}(1991)}]{Allen:1991}
\bibinfo{author}{\bibfnamefont{S.~J.} \bibnamefont{Allen Jr.~\emph{et al.}}},
  \bibinfo{journal}{Phys. Rev. B} \textbf{\bibinfo{volume}{43}},
  \bibinfo{pages}{9599} (\bibinfo{year}{1991}).

\bibitem[{\citenamefont{Bogaerts~\emph{et al.}}(1992)}]{Bogaerts:1992}
\bibinfo{author}{\bibfnamefont{R.}~\bibnamefont{Bogaerts~\emph{et al.}}},
  \bibinfo{journal}{Physica B} \textbf{\bibinfo{volume}{177}},
  \bibinfo{pages}{425} (\bibinfo{year}{1992}).

\bibitem[{\citenamefont{Bogaerts~\emph{et al.}}(1993)}]{Bogaerts:1993}
\bibinfo{author}{\bibfnamefont{R.}~\bibnamefont{Bogaerts~\emph{et al.}}},
  \bibinfo{journal}{Physica B} \textbf{\bibinfo{volume}{184}},
  \bibinfo{pages}{232} (\bibinfo{year}{1993}).

\bibitem[{\citenamefont{Bogaerts\emph{et al.}}(1996)}]{Bogaerts_et_al:1996}
\bibinfo{author}{\bibfnamefont{R.}~\bibnamefont{Bogaerts\emph{et al.}}},
  \bibinfo{journal}{Phys. Rev. B} \textbf{\bibinfo{volume}{53}},
  \bibinfo{pages}{15951} (\bibinfo{year}{1996}).

\bibitem[{\citenamefont{Georges~\emph{et al.}}(1996)}]{geo96}
\bibinfo{author}{\bibfnamefont{A.}~\bibnamefont{Georges~\emph{et al.}}},
  \bibinfo{journal}{Rev. Mod. Phys.} \textbf{\bibinfo{volume}{68}},
  \bibinfo{pages}{13} (\bibinfo{year}{1996}).

\bibitem[{\citenamefont{Vollhardt~\emph{et al.}}(2005)}]{voll05}
\bibinfo{author}{\bibfnamefont{D.}~\bibnamefont{Vollhardt~\emph{et al.}}},
  \bibinfo{journal}{J. Phys. Soc. Jpn.} \textbf{\bibinfo{volume}{74}},
  \bibinfo{pages}{136} (\bibinfo{year}{2005}).

\bibitem[{\citenamefont{Kotliar~\emph{et al.}}(2006)}]{kotliar_review}
\bibinfo{author}{\bibfnamefont{G.}~\bibnamefont{Kotliar~\emph{et al.}}},
  \bibinfo{journal}{Rev. Mod. Phys.} \textbf{\bibinfo{volume}{76}},
  \bibinfo{pages}{865} (\bibinfo{year}{2006}).

\bibitem[{\citenamefont{Hubbard}(1963)}]{hubbard_1}
\bibinfo{author}{\bibfnamefont{J.}~\bibnamefont{Hubbard}},
  \bibinfo{journal}{Proc. Roy. Soc. (London)} \textbf{\bibinfo{volume}{A 276}},
  \bibinfo{pages}{238} (\bibinfo{year}{1963}).

\bibitem[{\citenamefont{Pourovskii~\emph{et al.}}(2007)}]{SCDMFT}
\bibinfo{author}{\bibfnamefont{L.~V.} \bibnamefont{Pourovskii~\emph{et al.}}},
  \bibinfo{journal}{Phys. Rev. B} \textbf{\bibinfo{volume}{76}},
  \bibinfo{pages}{235101} (\bibinfo{year}{2007}).

\bibitem[{\citenamefont{Carnal~\emph{et al.}}(1989)}]{Carnall89}
\bibinfo{author}{\bibfnamefont{W.~T.} \bibnamefont{Carnal~\emph{et al.}}},
  \bibinfo{journal}{J.~Chem. Phys.} \textbf{\bibinfo{volume}{90}},
  \bibinfo{pages}{3443} (\bibinfo{year}{1989}).

\bibitem[{\citenamefont{Lea et~al.}(1962)\citenamefont{Lea, Leask, and
  Wolf}}]{Lea1962}
\bibinfo{author}{\bibfnamefont{K.~R.} \bibnamefont{Lea}},
  \bibinfo{author}{\bibfnamefont{J.~M.} \bibnamefont{Leask}}, \bibnamefont{and}
  \bibinfo{author}{\bibfnamefont{W.~P.} \bibnamefont{Wolf}},
  \bibinfo{journal}{J.~Phys. Chem. Solids} \textbf{\bibinfo{volume}{23}},
  \bibinfo{pages}{1381} (\bibinfo{year}{1962}).

\bibitem[{\citenamefont{Allen Jr.~\emph{et al.}}(1989)}]{Allen_et_al:1989}
\bibinfo{author}{\bibfnamefont{S.~J.} \bibnamefont{Allen Jr.~\emph{et al.}}},
  \bibinfo{journal}{Phys. Rev. Lett.} \textbf{\bibinfo{volume}{62}},
  \bibinfo{pages}{2309} (\bibinfo{year}{1989}).

\bibitem[{\citenamefont{Anisimov~\emph{et
  al.}}(1997)}]{anisimov_lda+u_review_1997_jpcm}
\bibinfo{author}{\bibfnamefont{V.~I.} \bibnamefont{Anisimov~\emph{et al.}}},
  \bibinfo{journal}{J. Phys. Condensed Matter} \textbf{\bibinfo{volume}{9}},
  \bibinfo{pages}{767} (\bibinfo{year}{1997}).

\bibitem[{\citenamefont{Larson~\emph{et al.}}(2007)}]{Larson2007}
\bibinfo{author}{\bibfnamefont{P.}~\bibnamefont{Larson~\emph{et al.}}},
  \bibinfo{journal}{Phys. Rev. B} \textbf{\bibinfo{volume}{75}},
  \bibinfo{pages}{045114} (\bibinfo{year}{2007}).

\bibitem[{\citenamefont{Komesu~\emph{et al.}}(2003)}]{Komesu_et_al:2003}
\bibinfo{author}{\bibfnamefont{T.}~\bibnamefont{Komesu~\emph{et al.}}},
  \bibinfo{journal}{Phys. Rev. B} \textbf{\bibinfo{volume}{67}},
  \bibinfo{pages}{035104} (\bibinfo{year}{2003}).

\bibitem[{\citenamefont{Duan~\emph{et al.}}(2004)}]{Duan2004}
\bibinfo{author}{\bibfnamefont{C.~G.} \bibnamefont{Duan~\emph{et al.}}},
  \bibinfo{journal}{Surface Review and Letters} \textbf{\bibinfo{volume}{11}},
  \bibinfo{pages}{531} (\bibinfo{year}{2004}).

\bibitem[{\citenamefont{Shim et~al.}(2007)\citenamefont{Shim, Haule, and
  Kotliar}}]{shim07}
\bibinfo{author}{\bibfnamefont{J.~H.} \bibnamefont{Shim}},
  \bibinfo{author}{\bibfnamefont{K.}~\bibnamefont{Haule}}, \bibnamefont{and}
  \bibinfo{author}{\bibfnamefont{G.}~\bibnamefont{Kotliar}},
  \bibinfo{journal}{Nature} \textbf{\bibinfo{volume}{446}},
  \bibinfo{pages}{513} (\bibinfo{year}{2007}).

\bibitem[{\citenamefont{Lambrecht~\emph{et al.}}(1997)}]{Lambrecht_et_al:1997}
\bibinfo{author}{\bibfnamefont{W.~R.~L.} \bibnamefont{Lambrecht~\emph{et
  al.}}}, \bibinfo{journal}{Phys. Rev. B} \textbf{\bibinfo{volume}{55}},
  \bibinfo{pages}{9239} (\bibinfo{year}{1997}).

\bibitem[{\citenamefont{Chantis et~al.}(2007)\citenamefont{Chantis, van
  Schilfgaarde, and Kotani}}]{chantis07}
\bibinfo{author}{\bibfnamefont{A.~N.} \bibnamefont{Chantis}},
  \bibinfo{author}{\bibfnamefont{M.}~\bibnamefont{van Schilfgaarde}},
  \bibnamefont{and} \bibinfo{author}{\bibfnamefont{T.}~\bibnamefont{Kotani}},
  \bibinfo{journal}{Phys. Rev. B} \textbf{\bibinfo{volume}{76}},
  \bibinfo{pages}{165126} (\bibinfo{year}{2007}).

\bibitem[{\citenamefont{Wulff~\emph{et al.}}(1988)}]{Wulff88}
\bibinfo{author}{\bibfnamefont{M.}~\bibnamefont{Wulff~\emph{et al.}}},
  \bibinfo{journal}{Europhys. Lett.} \textbf{\bibinfo{volume}{7}},
  \bibinfo{pages}{629} (\bibinfo{year}{1988}).

\end{thebibliography}

\end{document}